\documentclass[aps,pre,superscriptaddress,showpacs]{revtex4}
\usepackage[utf8]{inputenc}
\usepackage{amsmath}
\usepackage{amsfonts}
\usepackage{graphicx}

\begin{document}

\title{One-dimensional long-range percolation: a numerical study}

\author{G. Gori}
\email[Corresponding author: ]{gori@sissa.it}
\affiliation{CNR-IOM DEMOCRITOS Simulation Center, Via Bonomea 265, I-34136 Trieste, Italy}

\author{M. Michelangeli}
\affiliation{SISSA, Via Bonomea 265, I-34136 Trieste, Italy}

\author{N. Defenu}
\affiliation{SISSA, Via Bonomea 265, I-34136 Trieste, Italy}
\affiliation{CNR-IOM DEMOCRITOS Simulation Center, Via Bonomea 265, I-34136 Trieste, Italy}

\author{A. Trombettoni}
\affiliation{CNR-IOM DEMOCRITOS Simulation Center, Via Bonomea 265, I-34136 Trieste, Italy}
\affiliation{SISSA, Via Bonomea 265, I-34136 Trieste, Italy}
\affiliation{INFN, Sezione di Trieste, Via Bonomea 265, I-34136 Trieste, Italy}

\begin{abstract}
In this paper we study bond percolation on a one-dimensional chain 
with power-law bond probability $C/ r^{1+\sigma}$, where $r$ is the distance
length between distinct sites.
We introduce and test an order $N$ Monte Carlo 
algorithm and we determine as a function of $\sigma$ 
the critical value $C_{c}$ 
at which percolation occurs.
The critical exponents in the range $0<\sigma<1$ are reported 
and compared with mean-field and $\varepsilon$-expansion results. 
Our analysis is in agreement, up to a numerical precision $\approx 
10^{-3}$, with the mean field result for the anomalous 
dimension $\eta=2-\sigma$, showing 
that there is no correction to $\eta$ due to correlation effects. 
\end{abstract}

\maketitle

\section{Introduction}

Percolation is a paradigmatic model of statistical mechanics
\cite{broadbent_1957}. Despite its simplicity it is able 
to capture many features of real world phenomena
ranging from coffee brewing to oil industry (see \cite{grassberger_2014}
for a recent review) and at the same 
time it allows for various exact results \cite{grimmett_1999,bollobas_2006}.
One of the most remarkable features of percolation 
is to display phase transition phenomena with nontrivial
geometric properties \cite{essam_1980} 
prototyping in a simplified and essential 
stripped-to-the-bone setting features found in more
sofisticated models.

A significant bulk of the work regarding
percolation deals with models 
where probability of connecting 
two elements is zero beyond
a given range, i.e. short-range (SR)
models. The typical SR percolation refers to a non-vanishing probability 
to connect nearest neighbour elements or sites of a lattice 
\cite{grimmett_1999,bollobas_2006}. 
In SR percolation the one-dimensional limit 
is trivial: if one denotes by $C$ 
the probability to connect with a bond 
two nearest neighbour sites one immediately sees that in one dimension 
the critical value $C_c$ equals $1$ (otherwise the two ends of the chain cannot 
be connected). This result parallels the well known results that for 
classical Ising and \verb|O(N)|  
chains the critical temperature is vanishing for SR interactions 
\cite{mussardo_2010,ortiz_2011}. However it is well known 
that adding long-range (LR) couplings in the one-dimensional 
Ising model one may have a non vanishing critical 
temperature \cite{thouless_1969,dyson_1971}. Therefore 
in the literature it has been explored 
the percolation in presence of long-range (LR) interactions, 
for which one expects a percolative transition also in one dimension 
\cite{aizenmann_1986}. 

Systems where the elementary constituents
affect each other with laws that decay
slowly as a function of their mutual distance
play a central role in physics including, among
others, gravitational systems, unscreened
plasmas and (di)polar systems \cite{campa_2014}. 
A typical form of the interaction is $\propto 1/r^{d+\sigma}$, 
where $r$ is the distance between elements or sites of the system 
having dimension $d$, and $\sigma$ quantifies the range 
of the interactions.

The inclusion of LR interactions into 
percolation models, with bond probability of the 
form $\propto 1/r^{d+\sigma}$, 
proves valuable also for 
modeling of several phenomena
belonging to the wide arena of
complex systems, such as disease 
spreading \cite{grassberger_1983}, social aggregation \cite{solomon_2000} 
and financial transactions \cite{stauffer_2001}. A very interesting 
example of such studies is provided 
by epidemic spreading processes \cite{newman_2002}. In
the presence of SR infection, when no cooperative
effects are considered and there is immunization or death 
after the infection, the epidemic process gives rise 
to ordinary percolation clusters \cite{mollison_1977}. 
The generalization to cases where the dynamical process takes 
place in lattices where at least some of the infections are LR 
was considered \cite{biskup_2004,emmerich_2012,grassberger_2013b}. 
Further interest in the study of LR percolation 
was also triggered by papers 
giving some more exact results and its realization 
on finite graphs \cite{benjamini_2001,coppersmith_2002}.

The consideration of LR systems presents in general many interesting
features, with theoretical and numerical
challenges. Indeed in a renormalization
group (RG) setting the SR Wilson-Fisher
fixed point can be altered by
sufficiently strong LR interactions,  
leading at criticality to different 
universality classes. In particular
the effect of LR interactions 
on a system in $d$ dimensions
has been related (even though the mapping should not be exact) to a system
living in an effective dimension $d_{\mathrm{eff}} \geq d$.

The existence of such effective dimension relation between 
LR and SR systems comes from the low energy 
behavior of the critical propagator for LR interactions. 
Indeed using the conventional definition that 
the two point correlation functions at criticality scales as
$G(r)\sim r^{d-2+\eta}$, for a LR interacting system the anomalous
dimension correction can be rewritten as $\eta=2-\sigma+\delta\eta$, where
$2-\sigma$ is the result obtained in the mean-field approximation and 
$\delta\eta$ is the correction due to correlation effects. 

The expected anomalous dimension correction $\delta\eta$ 
was at the center of an long-lasting, intense 
debate. Classical early-day RG results on the Ising 
model conjectured $\delta\eta=0$ to be valid 
at all orders \cite{fisher_1972}. Successive investigations complemented 
this result introducing the threshold value $\sigma_{*}=2-\eta_{SR}$, where 
$\eta_{SR}$ is the anomalous dimension of a SR model in 
the same dimension of the LR one under study \cite{sak_1973}.  
At $\sigma=\sigma_{*}$ LR interactions become 
irrelevant with respect to SR ones \cite{sak_1973}. 
These results, in agreement with Monte Carlo findings \cite{luijten_2002}, 
have been questioned 
by numerical simulations on the two-dimensional LR Ising 
model indicating $\delta\eta\neq 0$ and $\sigma_{*}=2$ 
\cite{picco_2012,blanchard_2013}. 
We also observe that Monte Carlo simulations support 
the presence of SR behaviour into a finite amount of the region
$\sigma < 2$ for the $2D$ percolation \cite{linder_2008}, in agreement with the Sak's result 
\cite{sak_1973}. However recently presented results 
for susceptible-infected-removed epidemic processes with LR 
infection on a two-dimensional lattice 
appear to be in contraddiction with the Sak estimate 
for $\sigma_{*}$, 
pointing out to the possibility 
that at least some of the critical 
exponents are different for all $\sigma < 2$ from those of ordinary 
SR percolation \cite{grassberger_2013b}.  

These results eventually led to several numerical 
and analytic investigations over LR spin models. 
However up to date numerical simulations on the two-dimensional LR Ising 
model confirmed the traditional scenario with $\delta\eta=0$ 
and $\sigma_{*}=2-\eta_{SR}$ \cite{horita_2016}, 
also indicating an extremely slow convergence as a function of 
the system size for the critical amplitude of the standard Binder ratio 
of magnetization. Logarithmic corrections at the 
boundary value $\sigma_{*}$ are a possible 
source of error in numerical approaches.  
The occurrence of such logarithmic corrections at the threshold 
value $\sigma_{*}$ 
were confirmed by analytical results, which 
also indicated the results $\delta\eta=0$ and $\sigma_{*}=2-\eta_{SR}$ 
\cite{parisi_2014,defenu_2014}. Numerical results 
for the two-dimensional LR percolation were presented 
in \cite{parisi_2014}. The conformal invariance at criticality 
of the two-dimensional 
LR Ising model \cite{rychkov_2015} and in quantum LR chains 
\cite{lepori15,lepori16} has been as well recently addressed. 
Moreover using both scaling arguments 
\cite{angelini_2014} and functional 
RG techniques \cite{defenu_2014} the results for the
critical exponents of a LR interacting model in dimension $d$ 
with decay exponent $\sigma$ were related to the 
ones of a SR model in dimension $d_{\mathrm{eff}}=(2-\eta_{SR})d/\sigma$. 
Such effective dimension relation was discussed in 
\cite{angelini_2014}, finding very good agreement with numerical Monte Carlo 
results. One can also see that the effective dimension 
is valid only at low approximation level 
in the RG treatment, even though the error introduced by considering 
it exact has been quantified to be of $\lesssim 1\%$ for the critical 
exponent $\nu$ for the two-dimensional LR Ising models \cite{defenu_2014}. 
We observe that in one-dimensional LR Ising and percolation models 
Sak's results implies $\sigma_{*}=1$ (since $\eta_{SR}=1$). 
A Berezinskii-Kosterlitz-Thouless (BKT)  
transition may be expected at $\sigma=1$ for both one-dimensional 
Ising and percolation models, which has been deeeply investigated 
in Ising chains \cite{luijten_2001} and also observed for 
susceptible-infected-removed epidemic processes with LR 
infection in one dimension \cite{grassberger_2013a}.

Despite the wide appeal of the LR percolation 
model some important information
is still missing. Indeed, for the equilibrium properties, 
no estimate of the critical threshold
and of critical exponents
for different values of the decay 
parameter $\sigma$ is available to date, at the best of our knowledge, 
for the LR one-dimensional percolation.
The goal of the present paper is to give
numerical estimates of these
quantities. To this aim we develop a 
Monte Carlo algorithm allowig to efficiently simulate 
one-dimensional LR bond percolation. As a side result we can 
numerically check the validity of the Sak's prediction \cite{sak_1973} for the 
value of $\sigma_{*}$ in the one-dimensional LR percolation.
 
The paper is organized as follows: in 
Section \ref{model} we present
the model and briefly remind previous
results relevant for our work. In 
Section \ref{algorithm} we present
a novel order-$N$ algorithm for the study
of LR percolation models and discuss
its application to a LR 
version of Swendsen-Wang algorithm.
In Sections \ref{thresholds} and
\ref{exponents} we respectively present 
our findings for the critical thresholds
and the critical exponents.
In Section \ref{conclusions} conclusions
are drawn.

\section{The model} \label{model}

\begin{figure}
 \includegraphics[width=.6\textwidth]{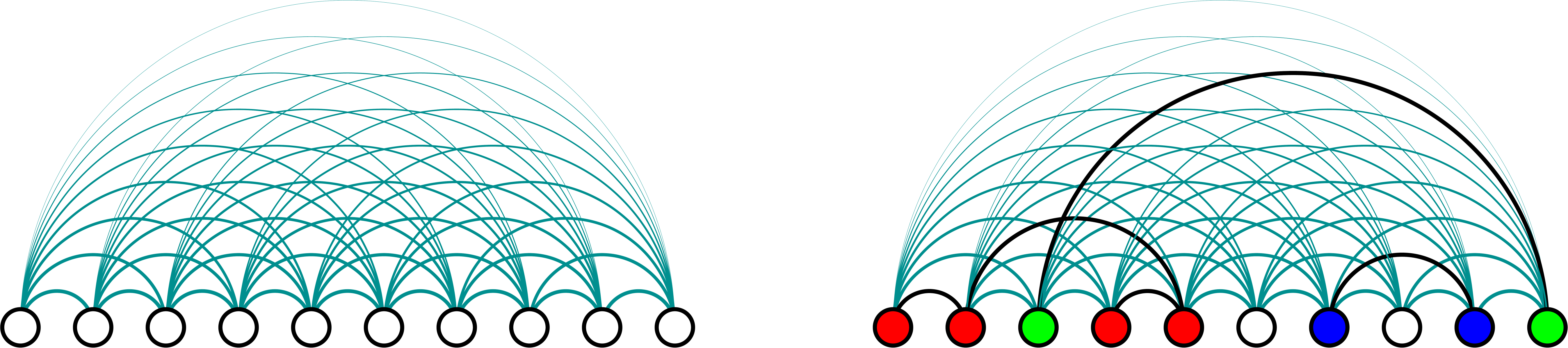}
 \caption{Pictorial representation of the LR bond percolation model. The lines
represent the possible bonds with width decreasing
as the connection probability decreases (left panel).
An instance of the model toghether with the
clusters generated (right panel).}
\label{figg1}
\end{figure}

We will consider a bond percolation
model on a one-dimensional lattice 
having $N$ sites with occupation probability $p_{i,j}$
of the bond connecting sites 
$i$ and $j$ given by (with $i \neq j$):
\begin{equation}
 p_{i,j}=p_{|i-j|}=\frac{C}{|i-j|^{1+\sigma}}
 \label{prob} 
\end{equation}
where $C\le 1$. See Figure \ref{figg1} for a graphical 
depiction of the model \eqref{prob}. When periodic boundary
conditions are used, an appropriate periodic distance
will be chosen, defined by the minimum
distance among periodic images of 
the sites $i$ and $j$. 

It is clear that if $\sigma \to \infty$ 
then the SR bond percolation with a bond probability 
$C$ between nearest neighbour sites is obtained. Notice 
that in the SR limit it is $C_c=1$. 

It is known that for the bond probability \eqref{prob} 
one has $C_c=1$ for $\sigma \ge 2$ \cite{aizenmann_1986}. Notice that if $p_{|i-j|}$ 
decays for large $|i-j|$ as $\propto \frac{1}{|i-j|^{1+\sigma}}$, 
then one has $C_c=1$ for $\sigma > 1$, but it is not necessarily 
$C_c=1$ for $\sigma=1$: in other words, the short distance 
behaviour of the probability \eqref{prob} may give rise to 
a percolation transition also at $\sigma=1$ \cite{aizenmann_1986}.

For future convenience we also mention the exact result by Schulman
\cite{schulman_1983}: comparing 
the LR percolation model
to a related percolation model on the Bethe lattice
a rigorous lower bound
for the critical threshold $C_c$ was obtained 
\begin{equation}
 C_c \geq C_c^{\rm Bethe} = \frac{1}{2 \zeta(1+\sigma)},\label{bethe_bound}
\end{equation}
where $\zeta(x)=\sum_{n=1}^{\infty} \frac{1}{n^{x}}$
is the Riemann zeta function.

\section{The algorithm}\label{algorithm}

The numerical study of percolation has largely
benefitted from efficient clustering
algorithms. Such algorithms are able to
track the cluster to which the
local degrees of freedom belong to. 
Since the inception of the
efficient Hoshen-Kopelman \cite{hoshen_1976} relabeling scheme
these algorithms have been brought closer 
to the optimal rigorous bounds predicted for union 
and find algorithm class to which they belong.

Simulation of LR models has always presented
major challenges on its own. The fact that when we 
perform a move/update of a system degree of freedom we 
have to deal with the state of all the other
degrees of freedom, unlike SR systems, 
typically brings the time for
a global update of the system from $O(N)$ to $O(N^2)$
severely reducing performance.

Recent advances in this field has been brought by
the introduction of algorithms scaling linearly
in the size of the system \cite{luijten_1995}. The efficiency of these
algorithms relies upon the fact that
only a small fraction of local degrees of freedom are actually interacting. 
Indeed even if every local variable in principle interacts 
with all the other degrees of freedom it effectively does
it only with a $O(N)$ fraction of them (in the case of non-frustrated
interacting systems this condition often translates in the
request for an extensive energy).

Similarly, in our LR percolation model, the number of connected
bonds, among the $N^2$ available, is of the order $O(N)$.
More specifically consider the class of bonds
of length $l$: the probability $\pi(n)$ of having $n$ 
connected bonds out of the $N$ available (with periodic
boundary conditions) is a binomial
\begin{equation}
 \pi_l(n) = \binom{N}{n} p_l^{n} (1-p_l)^{N-n}\label{binomial}
\end{equation}
which average to $N p_l$. If we sum all 
these probabilities we get a total number of
connected links, in the thermodynamic limit, of
$N 2 C \zeta(1+\sigma)$ for our model.
The above consideration suggests
the structure of an efficient $O(N)$ algorithm that is \cite{michelangeli_2015}:
\begin{itemize}
 \item extract, for each bond length $l$ class, the connected bonds 
according to \eqref{binomial}. This can be done in a rejection-free 
fashion by sampling the skip $s_l$ among two connected bonds which turn out to 
be distributed geometrically $p(s_l)\propto (1-p_l)^{s_l}$. The number of random numbers to be extracted is on average $1+p_l$
\item turn on the selected bond and cluster the sites according to the new connections. This can be done in constant time.
\end{itemize}
Note that the speedup can be traced back to 
having large classes of bonds with the same probability,
which is the case for translational invariant systems
(also with open boundary conditions).
The above algorithm, albeit its simplicity,
has not been discussed nor applied previously 
in the literature. Another remark is in order:
the rejection free extraction of connected bonds plays a central role, since a more naive 
sampling (i.e. scanning all the elements) would 
spoil the overall order-$N$ efficiency.

Interestingly this algorithm for LR bond percolation lends itself to
effectively implement the Swendsen-Wang algorithm \cite{swendsen_1987} 
in LR Ising or similar models. Indeed in the Swendsen-Wang algorithm
\cite{swendsen_1987} the cluster construction 
can be dealt with the above scheme. In contrast to what is usually done
one should \emph{first} extract a number of
tentatively connected bonds of length $l$
via our algorithm with the appropriate 
probability, that is $p_l = 1-e^{-\beta J_l}$ where
$J_l$ is the interaction among spins $l$ 
site apart, and \emph{later} 
connect them just if they have the same sign.
The cluster flipping part would be 
not altered. This provides an $O(N)$ algorithm
for LR Ising models.
Although LR adaptations of Swendsen-Wang algorithms 
scaling as $O(N \log N)$ \cite{luijten_1995}
and also $O(N)$ \cite{fukui_2009}
have already appeared in the literature, 
we believe our formulation
to be very coincise and clear since the mentioned
algorithms rely on binary search algorithms
and Walker alias method respectively which, albeit well
established, are more complicated to implement. 
Please notice however that our algorithm is well suited only for
translational invariant systems, while other 
available algorithms, mentioned before, apply also to non 
translational invariant systems.

\section{Observables}\label{observables}

Our estimates for the critical thresholds and critical 
exponents are based on a finite size scaling analysis of two quantities: 
where possible 
the analysis is done with both.

The first observable we have monitored consists
of the average cluster size $S$:
\begin{equation}
 S=\left\langle 
\frac
{\sum_{\mathcal{C}}\#_\mathcal{C}^2}
{N} \right\rangle.
\end{equation}
This quantity is the analogue of the susceptibility of an Ising model
in the paramagnetic phase. Indeed it is expected to scale as 
\begin{equation}
 S\propto N^{\frac{\gamma}{\nu}}
\label{s_scaling}
\end{equation}
at the critical point \cite{luijten_2002}. 
Since according to scaling relations, $\gamma/\nu=2-\eta$ 
and $\eta=2-\sigma+\delta\eta$ for a LR system, 
the scaling of $S$ will be used to evaluate the 
correction $\delta\eta$ to the power law decay of the correlation functions 
at criticality.
In the following it will be shown that our results are in agreement with 
the widely accepted result $\delta\eta=0$, which,
as already mentioned, has been recently confirmed by extensive 
numerical simulations and theoretical investigations on the LR Ising model 
\cite{angelini_2014,defenu_2014,horita_2016}. 

We have also introduced and studied the ratio:
\begin{equation}
 Q_G \equiv \left\langle 
\frac
{\sum_{\mathcal{C}}\#_\mathcal{C}^4}
{\left(\sum_{\mathcal{C}}\#_\mathcal{C}^2\right)^2} \right\rangle
\label{cum_G}
\end{equation}
where $\#_\mathcal{C}$ is the size of cluster $\mathcal{C}$, the
sum runs over all clusters and the 
brackets denote average over the different realizations of 
percolation. $Q_G$ measures the spread in the cluster 
size since when they have a characteristic finite size (in the
non percolating phase) it tends to zero as we approach the
thermodynamic limit; for $C>C_C$, when a macroscopic
(percolating) cluster develops, the ratio $Q_G$ tends to one
thus behaving as the Binder cumulant \cite{binder_1981}, 
that is widely used in magnetic systems.
if we had considered the ratio of cluster size
moments $\langle\sum_{\mathcal{C}}\#_\mathcal{C}^4\rangle
/\langle\left(\sum_{\mathcal{C}}\#_\mathcal{C}^2\right)^2\rangle$
as in \cite{deng_2007} we would have not observed 
any crossing \cite{michelangeli_2015}, as insted it occurs for $Q_G$.
Moreover since the value of $Q_G$ is ruled by the 
correlation length of the system we can put forward 
the following scaling form in the vicinity of $C_c$, for a 
finite system of size $N$:
\begin{equation}
 Q_G = Q_{G;c} + a_1 (C-C_c) N^{1/\nu} + a_2 (C-C_c) N^{2/\nu} +
     b_1 N^{\omega_1} + b_2 N^{\omega_2} + \ldots\label{qg_scaling}
\end{equation}

Our numerical investigations showed that from the crossing of $Q_G$ 
it is possible to have a good estimate of the critical thresholds 
$C_c(\sigma)$: to substantiate the use of $Q_G$ we report in Appendix 
\ref{cum} the results for $Q_G$ for the two-dimensional SR percolation 
on a square lattice, showing that the correct result can be retrieved. 
Regarding the critical exponents, it turns out that using $S$ gives good 
estimates for the critical exponents (we determined $\nu$ and $\eta$) for 
all the range $0<\sigma<1$, while $Q_G$ can be used for $\sigma$ close 
to $1$, where by the way it gives more precise estimates than the ones obtained 
by using $S$. Both methods, i.e. using $S$ and $Q_G$ give 
results in agreement between each others (where one obtains reliable findings). 
Finally, we observe that using $Q_G$ one can conclude that $C_c=1$ for 
$\sigma=1$ while using $S$ one numerically confirms that $C_c=0$ for 
$\sigma=0$, as it should be.

The runs have scanned different values of
$\sigma$, $N$ and $C$. For each value of the
parameters we used $4\cdot10^6$ different 
realizations. The considered size of the system
is up to $N=512\cdot10^3$. The random number generator
chosen is the MT19973 \cite{matsumoto_1998}.
The bulk of the experiments have been carried
out with open boundary conditions, 
while for selected values of the decay parameter
$\sigma$ we have also simulated periodic boundary
conditions.

\section{Critical thresholds}\label{thresholds}

\begin{figure}
 \includegraphics[width=.45\textwidth]{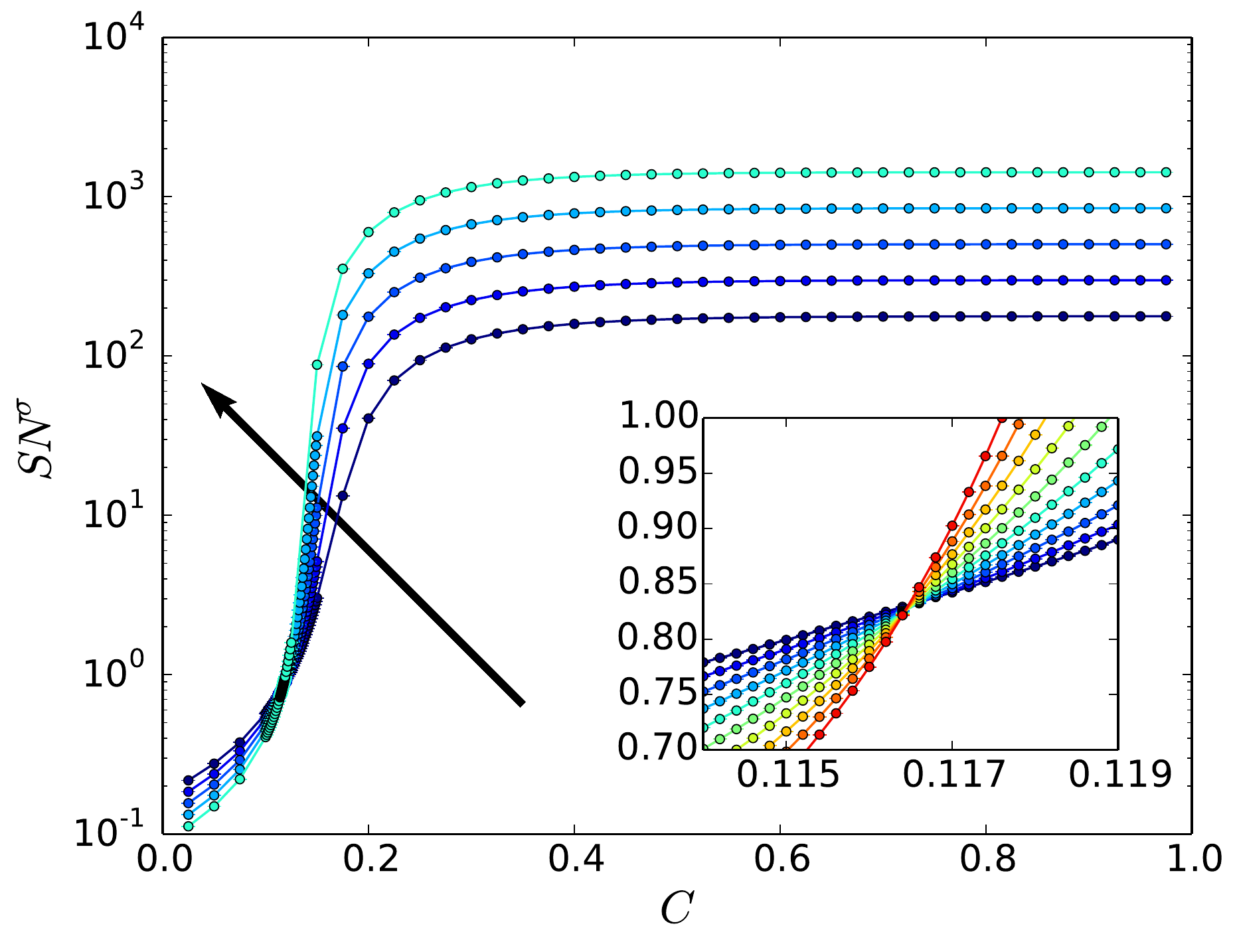}
 \includegraphics[width=.45\textwidth]{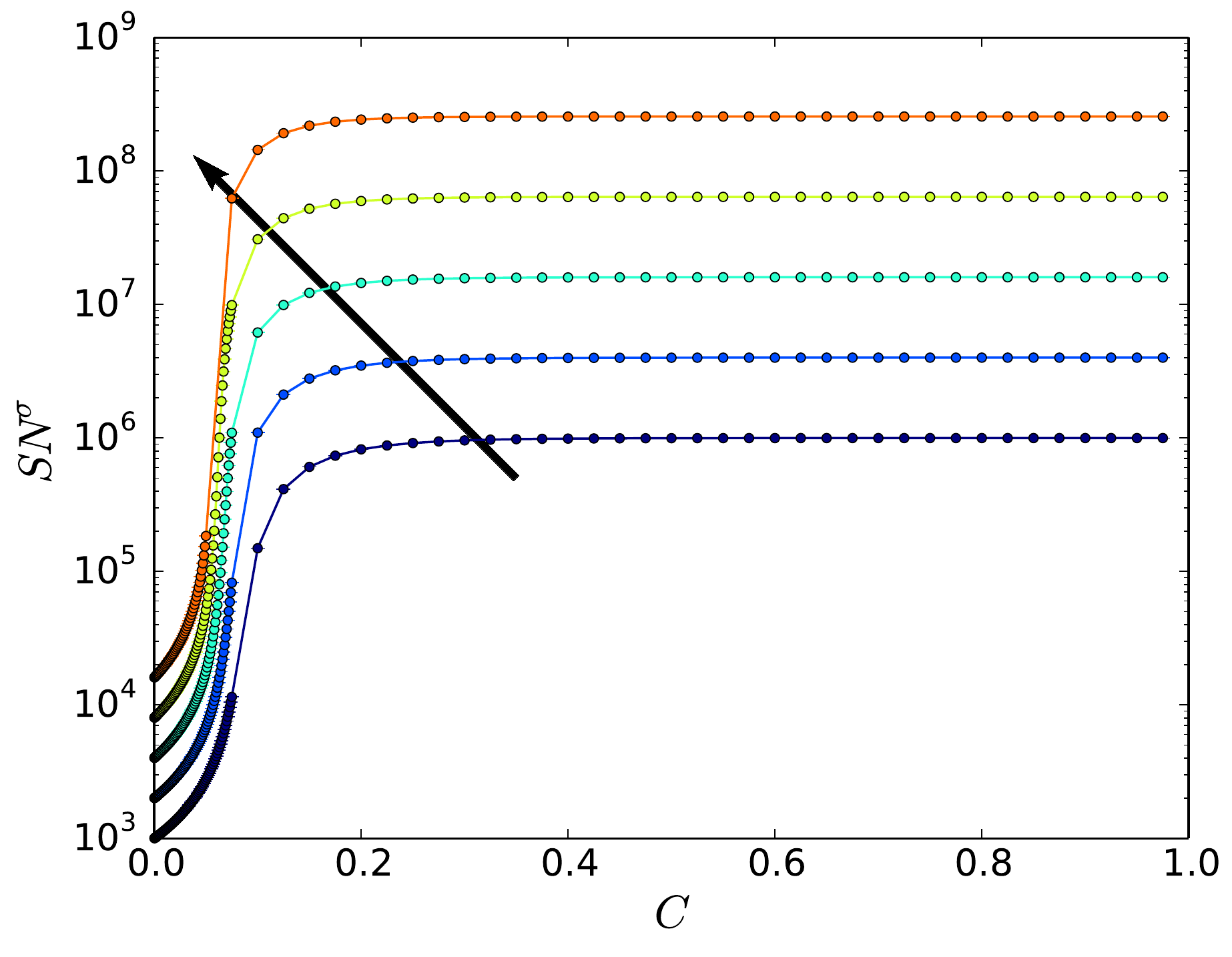}
 \caption{(Left panel) $S N^{\sigma}$ as a function of $C$
for $\sigma=0.25$. The lines refer to increasing (small) sizes 
$N=10^3$, $2\cdot10^3$, $4\cdot10^3$, $8\cdot10^3$, and 
$16\cdot10^3$ in the sense of the black arrow.
Errors are smaller than the size of the lines. In the inset a zoom 
of the figure around the crossing point is shown: the curves 
refer to sizes $N=10^3$, $2\cdot10^3$, $2^2\cdot10^3$, $2^3\cdot10^3$, 
$\ldots$, $2^9\cdot 10^3$ with the curves becoming
steeper as the system size is enlarged.  
(Right panel) Same as 
left panel with $\sigma=0$. As it is apparent, 
no crossing occurs.} 
 \label{fig2}
\end{figure}

To determine the critical thresholds $C_C$ as a function of $\sigma$ 
we study the quantity $S N^{\sigma}$
which 
should cross at $C_c$ (this is consistent with the fact that 
$\delta \eta=0$ and therefore $\gamma=\sigma \nu$, as it will be 
shown in the next Section). Indeed as one can see in 
Figure \ref{fig2} the curves for different sizes
meet at an almost $N$ independent point.

\begin{figure}
 \includegraphics[width=.45\textwidth]{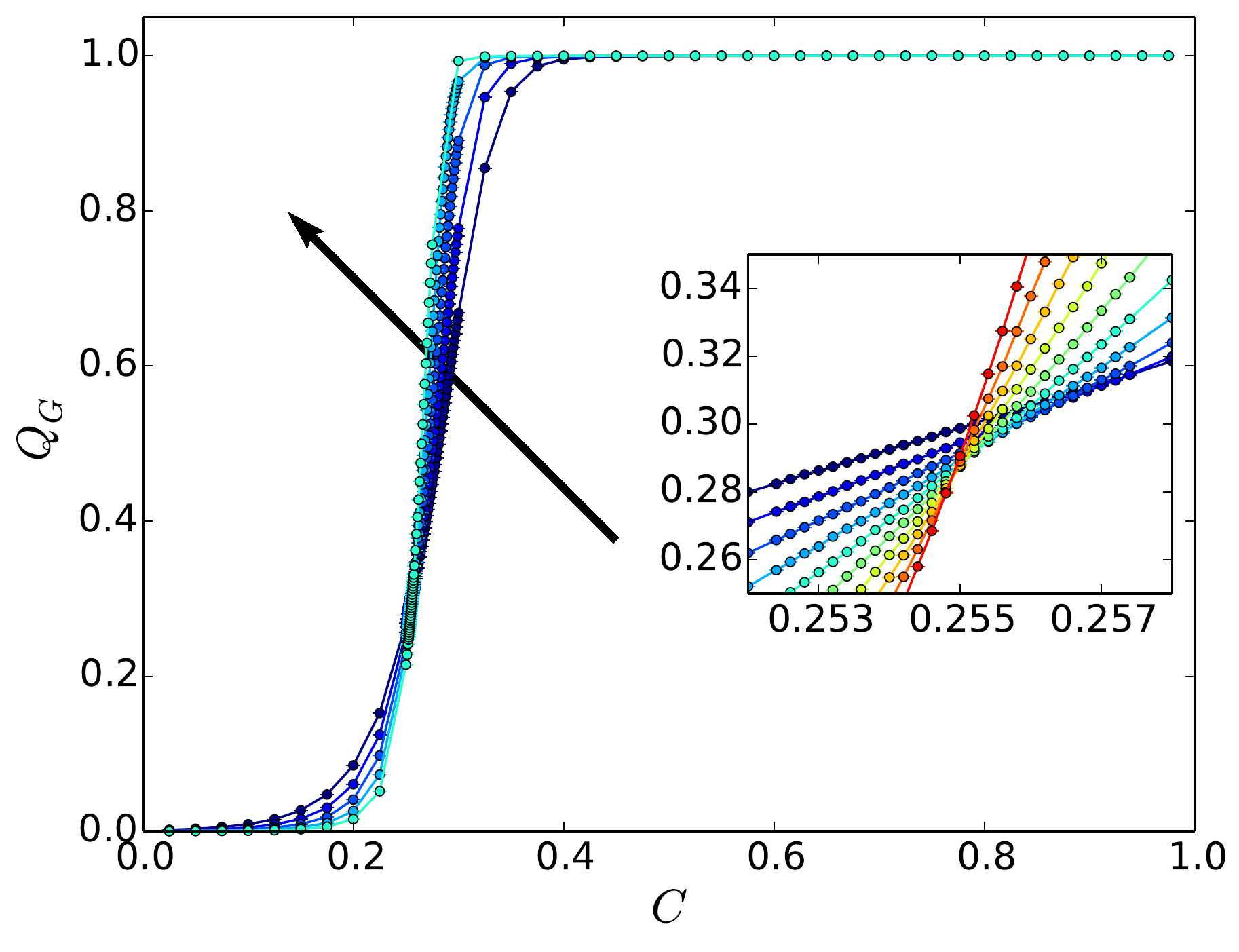}
 \includegraphics[width=.45\textwidth]{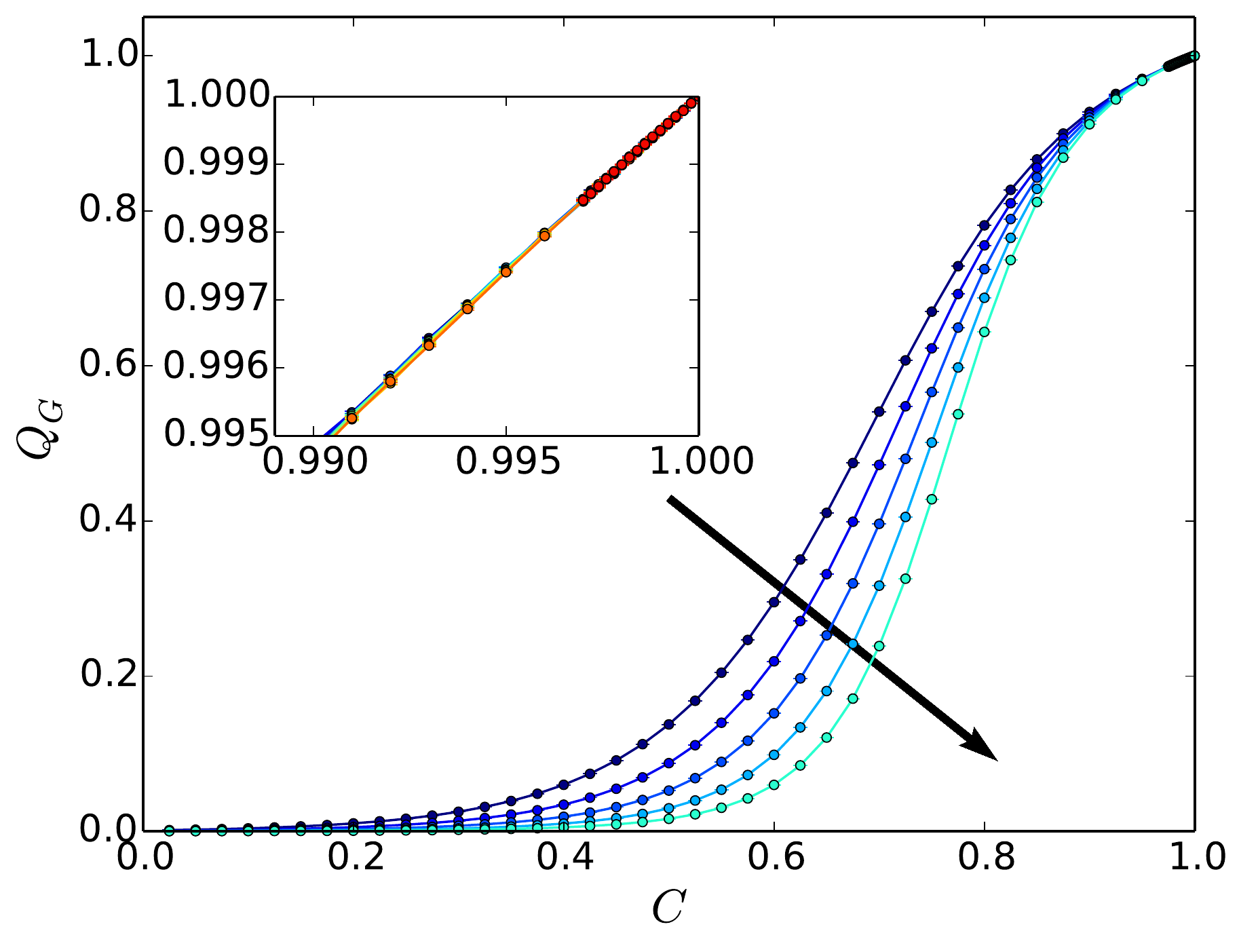}
 \caption{(Left panel) Ratio $Q$ as a function of $C$
for $\sigma=0.5$. The lines refer to increasing (small) sizes $N=10^3$,
$2\cdot10^3$, $4\cdot10^3$, $8\cdot10^3$, and $16\cdot10^3$ in the sense of the black arrow. Errors are smaller than the size of the lines. 
The inset shows the curves
in the vicinity of $C=1$ for larger system sizes (the same of the inset 
of Figure \ref{fig2}) as used to
locate the transition. (Right panel) Same as 
left panel with $\sigma=1$. In the inset a zoom 
around the $C=1$ point is reported: one can notice that the curves 
at the level of precision available do not show any crossing.}
 \label{fig3}
\end{figure}

Another way to locate the phase transition is to use 
the cumulant $Q_G$ defined in \eqref{cum_G}. 
The behavior of the ratio $Q_G$ as a function
of $C$ is depicted in the left panel of Figure \ref{fig3} for
the value of $\sigma=0.5$. Increasing the system size $Q_G$ approaches
a step profile jumping at 
the critical value of $C$ allowing for a
precise determination of the critical 
point $C_c$.
In the right panel of Figure \ref{fig3} we show the same plot
for $\sigma=1$. As we can see the curves,
within numerical error (see inset), do not cross for
$C<1$ supporting the theoretical prediction
that $C_c=1$ for the threshold value $\sigma=1$.

Both methods devised agree for most values
of $\sigma$ but the $Q_G$ crossing fails 
for low $\sigma$ because the crossing
happens in parts of the curve with large 
curvature (at the sizes considered). 
For our best estimates of $C_c$ we will
then rely on crossings of $S N^{\sigma}$.

\begin{figure}
 \includegraphics[width=.45\textwidth]{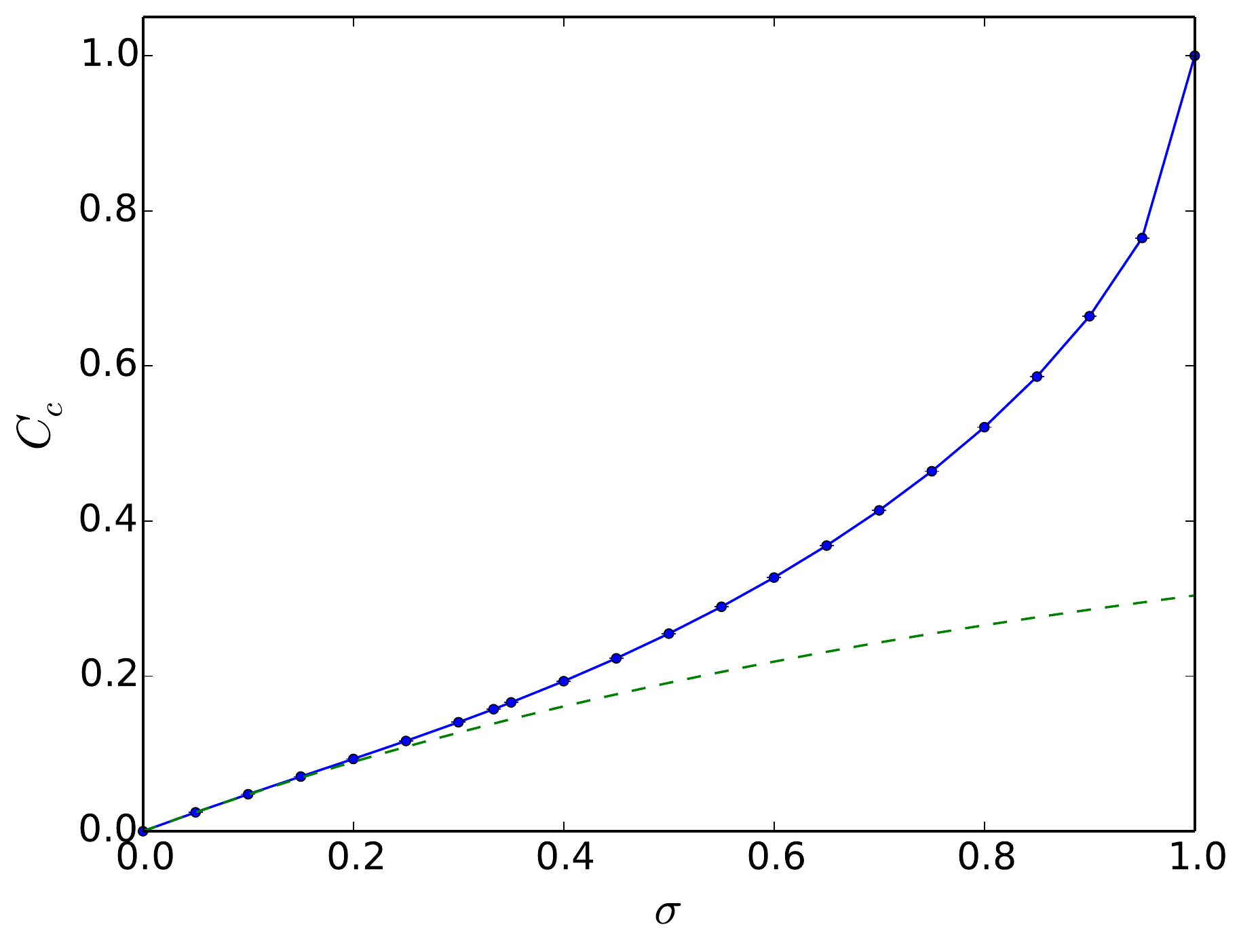}
 \caption{Critical thresholds $C_c$ as a function of
$\sigma$. The dotted line is the rigorous lower bound \eqref{bethe_bound}.}
 \label{fig4}
\end{figure}

The results of the above analysis yield the 
critical threshold as a function of $\sigma$ 
reported in the first column of Table \ref{tab1}
and for convenience in Figure \ref{fig4}
where we also report the Bethe lower bound \ref{bethe_bound}.
We notice that as $\sigma$ approaches
zero the estimated $C_c$ follows
$C_c^{\rm Bethe}$ closer and closer;
this is especially true in the
classical region $\sigma<d/3$. On the other
side as $\sigma\to 1$ the critical
threshold $C_c$ approaches the expected
value of $1$.

\begin{table}
\centering
\begin{tabular}{|l|l|l|l|}
  \hline
  $\sigma$ & $C_c$ & $\nu$ &  $\nu$($Q_G$-est.)\\ \hline
0.05 & 0.0243486(7) & 0.0497(2) & \qquad-\\
0.1 & 0.047685(8) & 0.1006(5) & \qquad-\\
0.15 & 0.070482(2) & 0.1503(3) & \qquad-\\
0.2 & 0.093211(16) & 0.2016(7) & \qquad-\\
0.25 & 0.11638(3) & 0.2474(7) & \qquad-\\
0.3 & 0.140546(17) & 0.305(8) & \qquad-\\
1/3 & 0.15736(6) & 0.340(4) & \qquad-\\
0.35 & 0.16610(5) & 0.344(3)& \qquad-\\
0.4 & 0.193471(15) & 0.355(9)& \qquad-\\
0.45 & 0.22293(6) & 0.363(12)& \qquad-\\
\hline
\end{tabular}
\begin{tabular}{|l|l|l|l|}
  \hline
  $\sigma$ & $C_c$ & $\nu$ & $\nu$($Q_G$-est.)\\ \hline
0.5 & 0.25482(5) & 0.3503(12) & \qquad-\\
0.55 & 0.289410(13) & 0.3532(13) & \qquad-\\
0.6 & 0.327098(6) & 0.3512(15) & \qquad-\\
0.65 & 0.368333(13) & 0.3495(5)& \qquad-\\
0.7 & 0.413752(14) & 0.341(2)& \qquad-\\
0.75 & 0.464202(19) & 0.3293(11) & \qquad-\\
0.8 & 0.521001(14) & 0.319(2) & 0.3161(4)\\
0.85 & 0.586264(10) & 0.299(2) & 0.2953(7)\\
0.9 & 0.66408(7) & 0.258(4) & 0.262(2) \\
0.95 & 0.76501(9) & 0.200(8) & 0.207(7)\\
  \hline
\end{tabular}
\caption{Estimated critical thresholds $C_c$ different values
of decay parameter $\sigma$. Values of the critical 
exponent $\nu$ are also reported. See text for details.}\label{tab1}
\end{table}

\section{Critical exponents}\label{exponents}

For the LR percolation model under exam we expect the existence of a 
SR effective dimension $d_{\mathrm{eff}}$ 
such that the universal quantities of the LR model can be related to
the ones of the corresponding SR in $d_{\mathrm{eff}}$ dimensions
\cite{ibanez_2013,angelini_2014,defenu_2014}. 
Such effective dimension relation reproduces standard scaling arguments 
and it reproduces correctly the qualitative features of phase diagrams \cite{defenu_2014}.
Also effective dimensional approach are very accurate close to the mean field region even
if generally not exact.

We have that for $\sigma<d/3$ the effective dimension relation 
is $d_{\mathrm{eff}}>6$ which is the mean-field region 
for the SR percolation problem \cite{amit_1976}. Moreover, 
in $d=1$ for $\sigma>1$ we have $d_{\mathrm{eff}}<2$ 
and the phase transition vanishes. These results are in 
agreement with the ones already present in LR percolation literature 
\cite{schulman_1983,zhang_1983}.

\begin{figure}[ht!]
 \includegraphics[width=.45\textwidth]{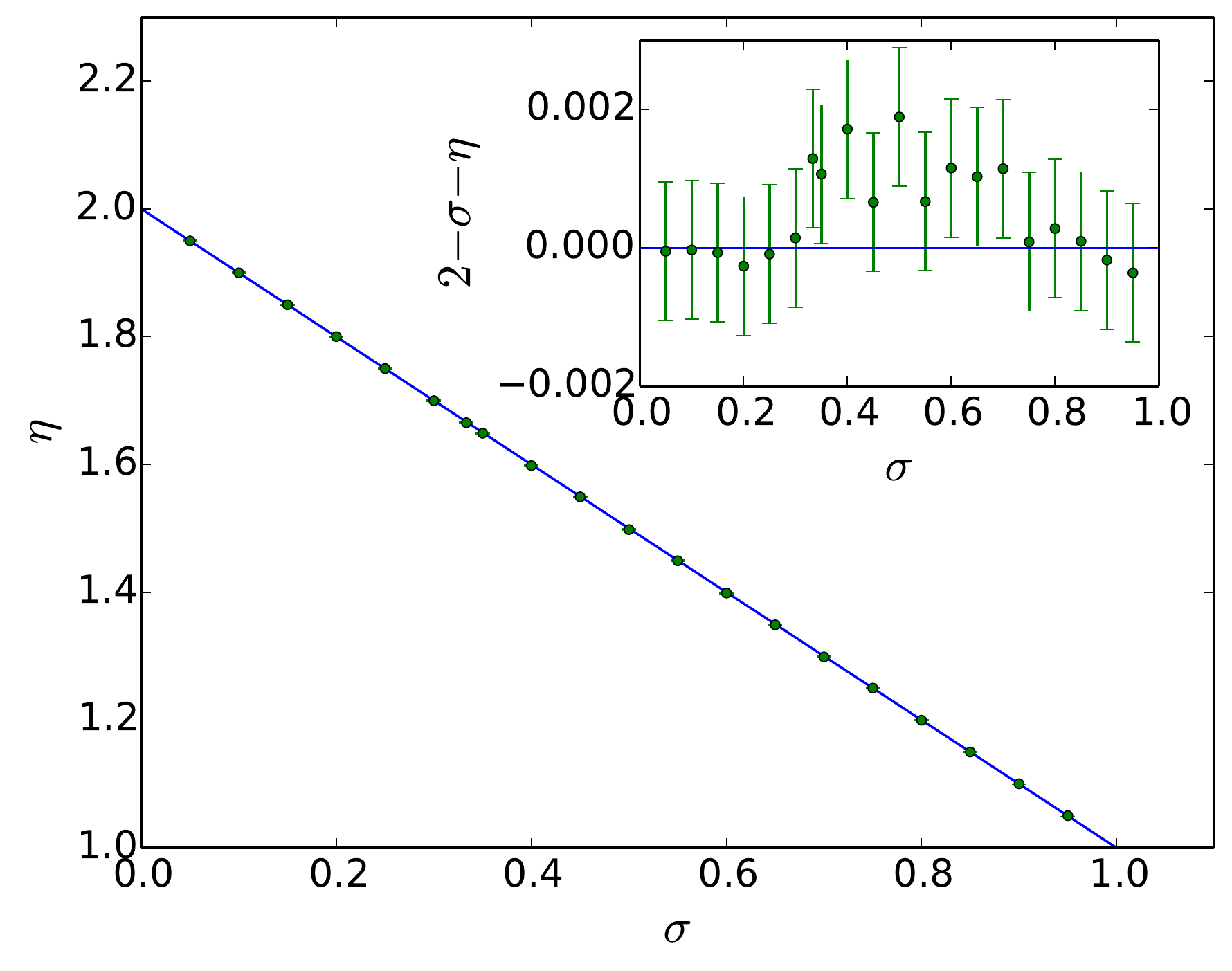}
 \caption{Estimated values for $\eta$ toghether with
predicted values from the RG prediction $\eta=2-\sigma$ 
\cite{sak_1973}. Inset: 
deviation from the RG prediction $\eta=2-\sigma$.}
 \label{fig5}
\end{figure}

\begin{figure}[ht!]
 \includegraphics[width=.45\textwidth]{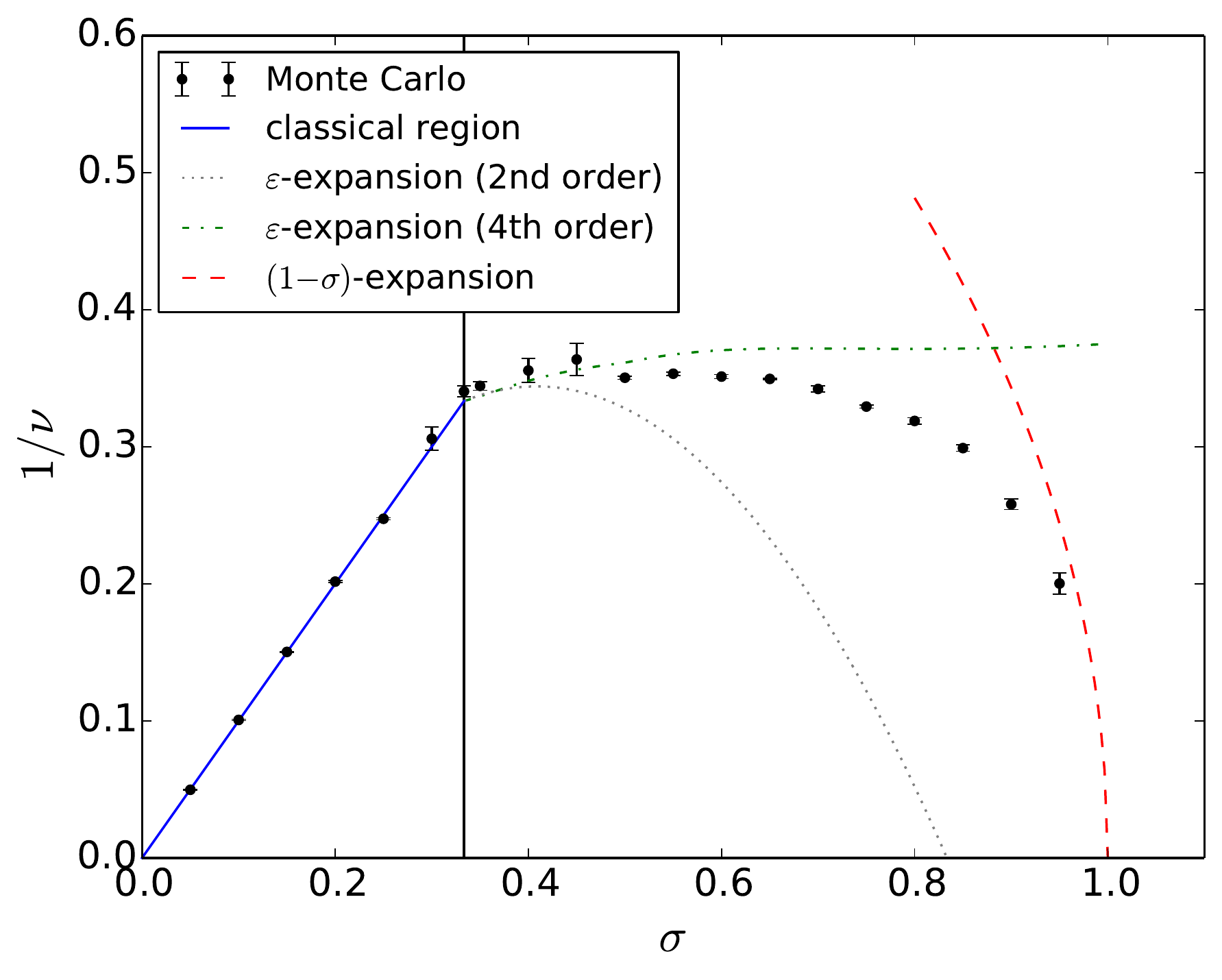}
 \caption{Estimated values for $1/\nu$ (circles) toghether with
predicted values from RG in the classical region ($\sigma<1/3$, blue
continuous line) and the results for the nonclassical region ($1/3<\sigma<1$); 
the red dashed line on the right of the figure is the prediction \eqref{bkt_pred}
(dashed red) valid close to the BKT point, while the gray and green lines 
are obtained the effective dimension \eqref{nu_rel} with the 
second (dotted gray, bottom) and fourth (dash-dotted green, top) $\varepsilon$-expansion 
results \eqref{vareps} taken from \cite{gracey_2015}.}
 \label{fig6}
\end{figure}

Our numerical results for $\delta \eta$ and $\nu$ as a function 
of $\sigma$ are reported in Figures \ref{fig5} and \ref{fig6} respectively. 
It is clear that the Sak prediction is rather clearly confirmed.
The values of $\delta\eta$ have been obtained by looking
for crossing of the curves $S N^{-\gamma/\nu}=S N^{-\sigma+\delta\eta}$ 
allowing for a nonzero value of $\delta\eta$. The best value 
has been chosen as the one giving the most constant value
of the crossing value $C$ for the largest sizes $N\geq16\cdot10^{3}$
considered. Since we found values of $\delta\eta$ consistent
with zero up to $1.5$ standard deviations we assumed $\eta=2-\sigma$
in the subsequent analysis.
Our best estimates of $\nu$ (reported in the third column of 
table \ref{tab1}) rely on the behaviour of $S N^{-\sigma}$
near $C_c$. Standard scaling arguments lead us
to expect the following scaling: $S N^{-\sigma}=\text{const.} + 
a_1 (C-C_c) N^{1/\nu} + a_2 (C-C_c) N^{2/\nu} + 
a_3 (C-C_c) N^{3/\nu} \ldots$; the analysis of the
data indeed confirms this expectation. Notice that in
contrast to the scaling for $Q_G$ (see \eqref{qg_scaling}) 
no subleading exponents $\omega_i$ were required
to explain the data. The region around the value
$\sigma=1/3$ requires a higher number of
powers of $(C-C_c)$ to be included in the fit
in order to keep $\chi_\text{d.o.f.}^2<1.5$ 
yielding a higher error on the estimates for $1/\nu$.
These data are given in the third column of
Table \ref{tab1}.
Finally an independent estimate of $1/\nu$
has been calculated via the more involved scaling \eqref{qg_scaling}
of $Q_G$ (reported in the fourth column of Table \ref{tab1}).
The values obtained are reliable only in a small
region around $0.8\leq \sigma <1$ but do confirm
the values retrieved with the observable $S$.

In order to benchmark our numerical results for $\nu$ it would be useful 
to have analytical predictions for it. 
However RG (and functional RG) calculations in LR percolation problems 
are less straightforward than the corresponding ones in LR Ising and 
\verb|O(N)| models \cite{defenu_2014}. The derivation of perturbative or 
non-perturbative approximated formulas 
goes beyond the scope of the present work. Nevertheless, 
it exists a shorthand to obtain approximated reference 
results to control the validity of our findings, 
relying on the effective dimension relation between LR and SR models. 
Indeed scaling arguments used to derive effective dimension relation 
$d_{\mathrm{eff}}$ are based on the field theory description and are thus 
 also valid for the percolation problem \cite{amit_1976}.

In a very recent paper \cite{gracey_2015} $\varepsilon$-expansion formulas 
on $\phi^{3}$ theory up to order $\varepsilon^4$ 
were derived and numerical estimates for the critical 
exponents in integer dimensions of SR percolation were given. 
We then can use the previously mentioned LR-SR effective dimension relation. 
According to scaling arguments and low order RG approximations 
the critical exponents of a LR model in dimension $d$ with exponent 
$\sigma$ are related to the ones of a SR model in dimension
\begin{equation}
 \label{effec_dim}
d_{\mathrm{eff}}=\frac{(2-\eta_{SR})d}{\sigma}.
\end{equation} 
The latter result was successfully applied 
in the case of two-dimensional LR Ising \cite{angelini_2014,defenu_2014} 
and LR \verb|O(N)| models \cite{defenu_2014}. In these works 
the correlation length exponent $\nu$ of the LR model 
was compared to the one of a SR model in dimension $d_{\mathrm{eff}}$ 
using the relation
\begin{equation}
\label{nu_rel}
\nu=\frac{(2-\eta_{SR})}{\sigma}\nu_{SR}(d_{\mathrm{eff}}).
\end{equation} 

However we expect our case to be rather different. Indeed in the 
two-dimensional case both the LR and SR systems are characterized by a 
spontaneous symmetry breaking of a $\boldsymbol{Z}_{2}$ symmetry. 
On the other hand the one-dimensional LR Ising model shows spontaneous 
symmetry breaking only for $\sigma<1$, while for $\sigma=1$ it undergoes 
a topological phase transition of the BKT type \cite{thouless_1969}, 
where kink excitations having logarithmic interactions \cite{cardy_1981}
play a role similar to 
the vortices in the standard SR XY model \cite{kosterlitz_1973}. 

The presence of BKT type phase transition is normally associated with 
a divergence of the correlation length exponent $\nu$ and then 
we expect the correlation length exponent of the $d=1$ LR Ising model 
to go to infinity as a function of $\sigma$ in the $\sigma\to 1$ limit, 
as it happens for the SR \verb|O(N)| models at $d=2$ in the 
$N\to 2$ limit \cite{codello_2015}. This result should not be reproduced by the 
effective dimension relation \eqref{nu_rel} since 
the SR Ising model should have finite correlation length exponent 
in any dimension. 

The case of LR percolation in one dimension is analogous 
to the case of the one-dimensional Ising model, indeed 
the percolation is the $q\to 1$ limit of the Potts model \cite{amit_1976}. 
Using the results of \cite{cardy_1981} in the limit $q \to 1$ one expects 
BKT behavior for LR percolation at $\sigma=1$. 
Then we expect our data to agree with the ones obtained from the 
SR results via relation \eqref{nu_rel} only far from $\sigma=1$ 
and close to the mean-field threshold $\sigma=\frac{1}{3}$. We can then 
discard anomalous dimension corrections in the SR models and use the 
simplified mapping
\begin{equation}
\label{nu_rel_simpl}
 \nu=\frac{2}{\sigma}\nu_{SR}(d_{\mathrm{eff}}).
\end{equation} 
with $d_{\mathrm{eff}}=\frac{2d}{\sigma}$ which is the $N\to \infty$ result 
\cite{defenu_2014}.
  
Therefore perturbative expressions for the critical exponents of 
LR percolation can be obtained imposing the relation 
between $d_{\mathrm{eff}}$ and $\varepsilon$ 
in the second order $\varepsilon$-expansion formulas 
obtained in \cite{essam_1980} and 
contained in \cite{gracey_2015} (notice that in \cite{gracey_2015} 
it is $d_{\mathrm{eff}} \equiv 6-2\varepsilon$). Defining 
$\tilde{\varepsilon}=\sigma-1/3$, valid in the one-dimensional system,
and expanding to second order in $\tilde{\varepsilon}$ we obtain
\begin{align}
\label{vareps}
\nu^{-1}&=
\frac{1}{3}+\frac{2}{7} \tilde{\varepsilon}-\frac{653}{343} 
\tilde{\varepsilon}^2+
O(\tilde{\varepsilon}^3)
\end{align}
which is plotted in Figure \ref{fig6} as a gray line.

Moreover in \cite{gracey_2015} results for the critical exponents are presented 
for integer dimensions using a Pad\'e approximant 
of $O(\varepsilon^4)$ expansion results constrained to reproduce the 
known exact values of the two-dimensional 
SR percolation. Interpolating such results a curve $\nu_{\mathrm{SR}}(d)$
for the correlation length exponent of the SR model as a function of 
the dimension
$d$ is obtained. This interpolation function is then transformed 
according to relation \eqref{nu_rel_simpl} with
$d_{\mathrm{eff}}=\frac{2\,d}{\sigma}$ yielding the results shown in 
Figure \ref{fig6} (green line). We observe that if we instead use the relations 
\eqref{effec_dim} and \eqref{nu_rel} we obtain practically the same results 
for $1/3 \lesssim \sigma \lesssim 0.6$ and worst results approaching 
$\sigma=1$ for the reasons explained above, while as expected 
poor results are obtained if non-resummed series 
for $\nu$ are used. 

A reasonable agreement is found with the numerical results 
even if it does not match with the numerical errors. 
This issue was expected due to the presence of logarithmic 
corrections at $\sigma=1/3$, which have not been included in 
our scaling assumptions and then lead to lack of accuracy close to 
the mean-field region. It is worth noting that logarithmic effects 
rapidly vanish since for $\sigma \gtrsim 0.4$ our results coincide 
with the RG ones, at least for $\sigma \lesssim0.5$ 
where perturbative expansion is expected to be still reliable.

Close to the boundary region $\sigma=1$ we can 
compare with the results of \cite{gracey_2015} 
which we generalized to the $\sigma\neq1$ case following 
the procedure of \cite{kosterlitz_1976}. 
Thus we obtain an analytic result for the correlation length exponent 
$\nu$ close to the BKT point $\sigma=1$:
\begin{equation}
\label{bkt_pred}
 \nu^{-1}=\sqrt{\frac{1-\sigma}{8q}}\left(q-2+\sqrt{4+12q+q^2+(q-2)\sqrt{2q(1-\sigma)}}\right),
\end{equation}
where the limit $q \to 1$ has to be taken.
 
Our results for the exponent $\nu$ are summarized in Figure \ref{fig6}.

\section{Conclusions}\label{conclusions}

In this paper we developed, tested and used an order-$N$ algorithm 
to study one-dimensional LR bond percolation. As a function 
of the power $\sigma$ of the decay of the bond probability we determined 
the critical threshold for percolation and the critical exponents 
$\eta$ and $\nu$. A precise estimate 
of the critical threshold is obtained by introducing 
a suitable geometric cumulant $Q_G$: 
comparison with exact results where available 
is reported. The value $\sigma_{*}$ above 
which no percolation occurs is retrieved to be $1$. 
For the critical exponent $\eta$ we 
compared our findings with the RG result $\eta=2-\sigma$ 
by Sak \cite{sak_1973}, getting 
agreement at the level of $10^{-3}$ precision. The exponent $\nu$ 
is compared with mean-field results and $\varepsilon$-expansion for the 
short-range percolation at an approximate effective dimension. For $\sigma<1/3$ 
agreement with the mean-field result for $\nu$ is found, while 
the $\varepsilon$-expansion provides a reasonable estimate 
in the region $1/3 \lesssim \sigma \lesssim 0.5$. Close to $\sigma=1$, 
where a Berezinskii-Kosterlitz-Thouless (BKT) transition is expected, an 
expansion in $1-\sigma$ has been derived.

An interesting future work would be to explore more in detail 
the region close to $\sigma=1$ and in particular to consider different 
short-distance behaviour of the bond probability with 
non trivial percolation thresholds $C_c$ at $\sigma=1$. The extension 
of the results presented in this paper to two-dimensional 
LR percolation may provide a worthwhile subject of future investigations: 
we expect that the critical exponents only depend on the ratio 
$d/\sigma$. 
Finally, 
we mention that with the algorithm discussed in this paper one can study 
the occurrence (or absence) of conformal invariance at criticality 
in two-dimensional LR percolation and the corresponding problem 
in one dimension.

\emph{Acknowledgement:} We gratefully acknowledge discussions with 
Miguel Ib\'anez Berganza, Luca Lepori, Stefano Ruffo, Slava Rychkov and 
Hirohiko Shimada. Hospitality in the Program 
"Conformal Field Theories and Renormalization Group Flows in 
Dimensions $d>2$" at the Galileo Galilei Institute for 
Theoretical Physics, Florence (Italy), where part of this work was performed, 
is gratefully acknowledged. GG and AT acknowledge funding support 
from "Progetto Premiale Anno 2012 ABNANOTECH - Atom-based technology".

\appendix

\section{The ratio $Q_G$ for short-range two-dimensional percolation}
\label{cum}

\begin{figure}
\includegraphics[width=.45\textwidth]{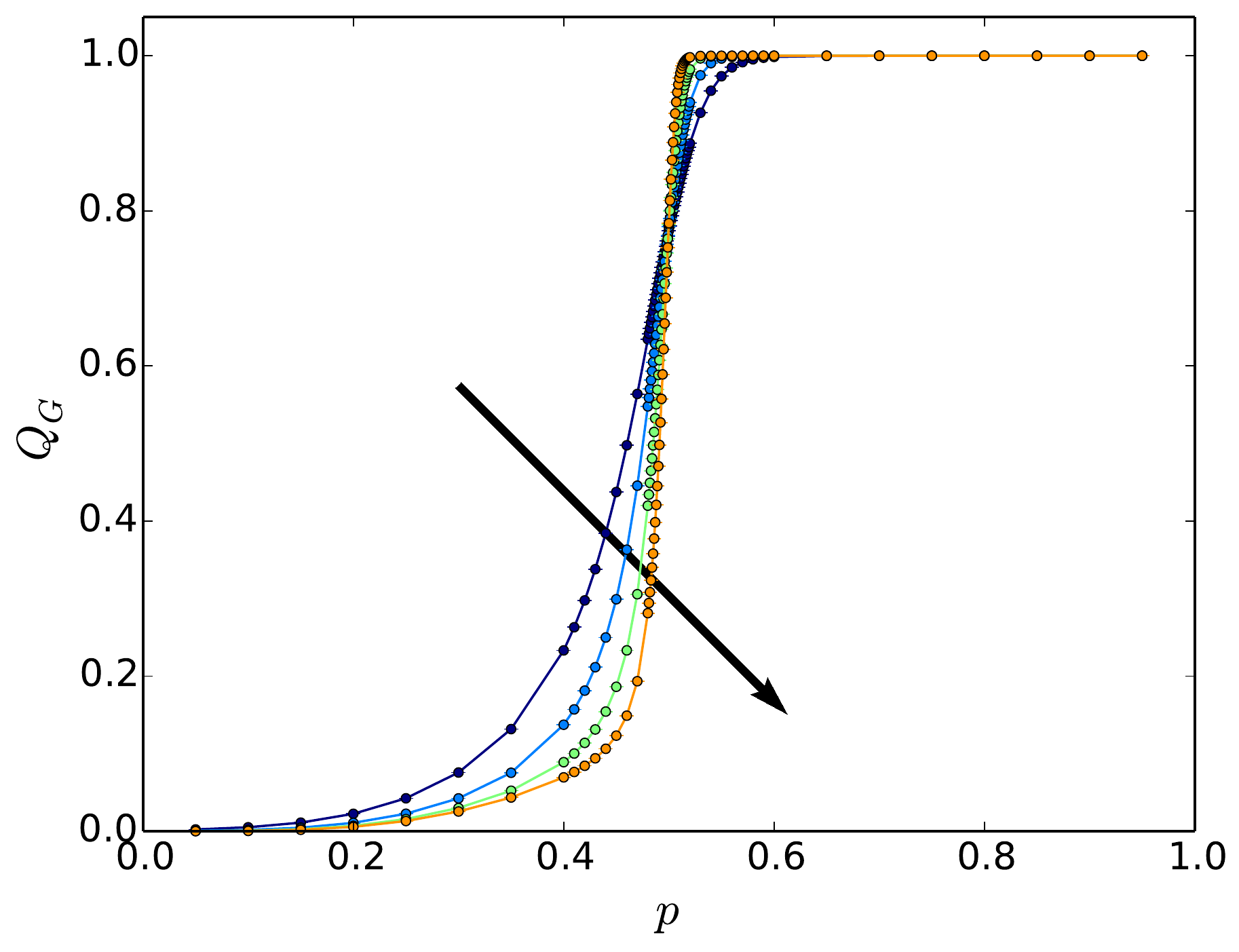}
\includegraphics[width=.45\textwidth]{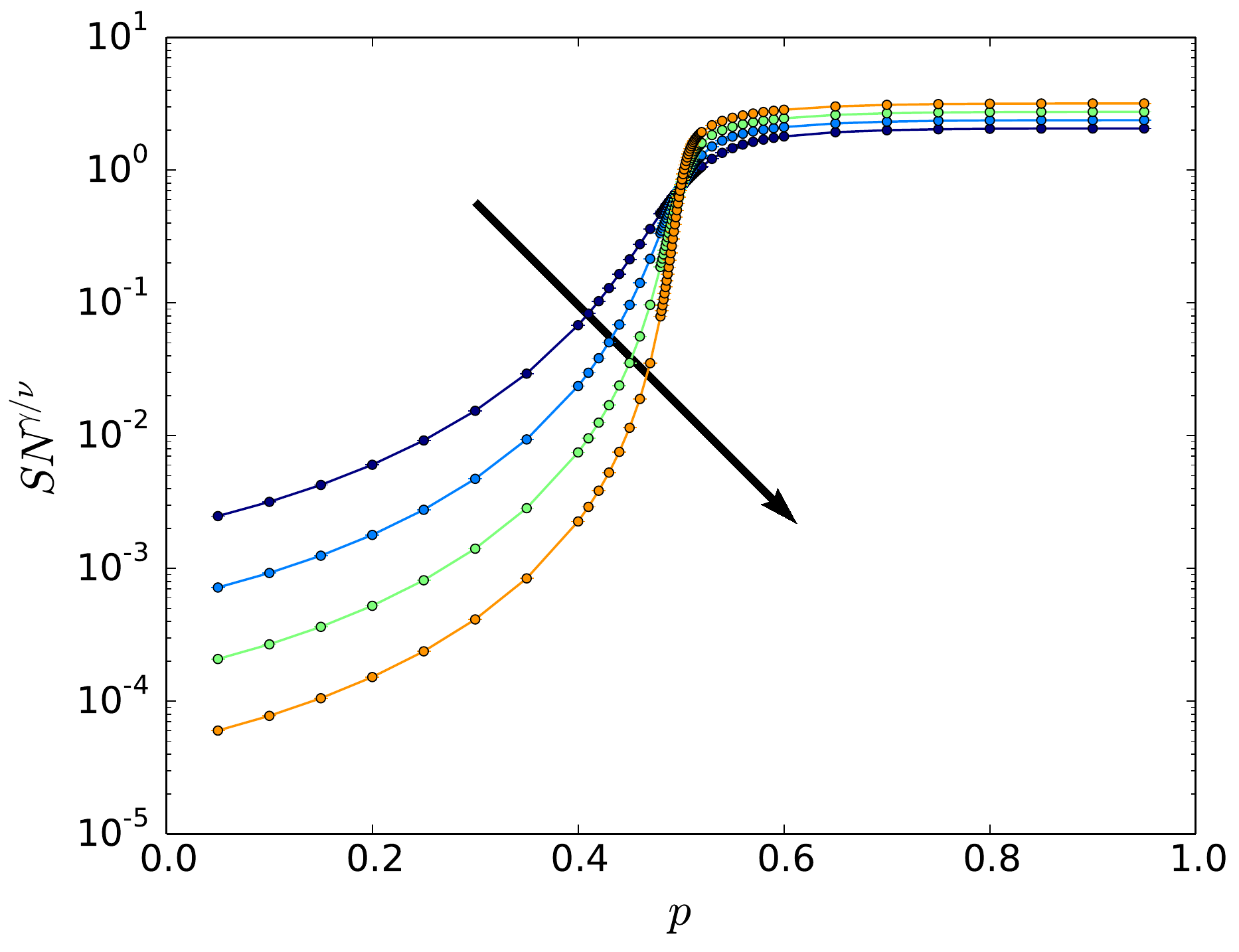}
 \caption{(Left panel) Ratio $Q_G$ as a function of $C$
for $\sigma=1/2$ for the SR bond percolation on a square lattice. 
The lines refer to increasing sizes $N=32$,
$64$, $128$, and $256$ in the sense of the black arrow.
Errors are smaller than the size of the lines. 
(Right panel) $S N^{\gamma/\nu}$ as a function of $p$
for the same sizes in the left panel.}
 \label{fig7}
\end{figure}

In order to assess the validity of the
ratio $Q_G$ introduced in Section \ref{observables}
for characterizing a percolation transition
we examine its behavior for a well
known percolation problem.
Bond percolation on a square lattice
is known to occur, due to symmetry
arguments, at an occupation probability
of $p=p_c=1/2$. In the left panel  of 
figure \ref{fig7}
we plot $Q_G$ for different system sizes.
As we can see the intersections approach the 
exact value. If we simply take the
best value for our estimate of
$p_c$ as the intersection of the
two largest systems considered
(i.e. $128\times 128$ and $256\times 256$)
we get $p_c=0.499997(4)$.
If we consider the crossing of the 
curves $S N^{-\gamma/\nu}$ instead, shown
in the right panel of figure \ref{fig7},
with the known value of $\gamma/\nu = 43/24$
we obtain the slightly less precise
estimate $p_c=0.4997(3)$.

\end{document}